\documentclass[aps,prl,twocolumn,superscriptaddress,showpacs,amsmath,
  nofootinbib]{revtex4} 
\usepackage{graphicx}
\usepackage{bm}
\usepackage{color}

\def\lsim{\mathrel{\rlap{\lower4pt\hbox{\hskip1pt$\sim$}}
    \raise1pt\hbox{$<$}}}         
\def\gsim{\mathrel{\rlap{\lower4pt\hbox{\hskip1pt$\sim$}}
    \raise1pt\hbox{$>$}}}         

\begin{document}


\title{Using the Standard Solar Model to Constrain Composition and S-Factors}



\author{Aldo Serenelli}
\email{aldos@ice.csic.es}
\affiliation{Instituto  de  Ciencias  del  Espacio  (CSIC-IEEC),  Facultad  de
  Ci\`{e}ncies, Campus UAB, 08193 Bellaterra, Spain}
\author{Carlos
  Pe\~{n}a-Garay} 
\email{penya@ific.uv.es}
\affiliation{Instituto de F\'{i}sica Corpuscular, CSIC-UVEG, Valencia 46071 Spain}
\author{W. C. Haxton}
\email{haxton@berkeley.edu}
\affiliation{Department  of Physics, University  of California,  Berkeley, and
  \\ Nuclear Science Division, Lawrence Berkeley National Laboratory,
Berkeley, CA 94720, USA}


\date{\today}

\begin{abstract}

While standard solar model (SSM)  predictions depend on approximately 20 input
parameters,  SSM neutrino  flux  predictions are  strongly  correlated with  a
single model output parameter,  the core temperature $T_c$.  Consequently, one
can extract physics from solar neutrino flux measurements while minimizing the
consequences of  SSM uncertainties, by  studying flux ratios  with appropriate
power-law weightings tuned to cancel  this $T_c$ dependence.  We re-examine an
idea for  constraining the  primordial C+N  content of the  solar core  from a
ratio of  CN-cycle $^{15}$O  to pp-chain $^8$B  neutrino fluxes,  showing that
nonnuclear SSM uncertainties  in the ratio are small  and effectively governed
by  a  single  parameter,  the  diffusion  coefficient.   We  point  out  that
measurements of  both CN-I  cycle neutrino branches  -- $^{15}$O  and $^{13}$N
$\beta$-decay --  could in  principle lead to  separate determinations  of the
core C  and N  abundances, due to  out-of-equilibrium CN-cycle burning  in the
cooler outer layers of the solar  core.  Finally, we show that the strategy of
constructing ``minimum  uncertainty" neutrino flux ratios can  also test other
properties of the SSM.  In particular, we demonstrate that a weighted ratio of
$^7$Be  and  $^8$B  fluxes constrains  a  product  of  S-factors to  the  same
precision currently possible with laboratory data.
\end{abstract}

\pacs{26.65.+t, 26.20.Cd,96.60.Fs, 14.60.Lm}

\maketitle


Important developments in  solar neutrino physics have occurred  over the past
one to two years that impact the  field's two major goals, probing the core of
the Sun and constraining  new weak-interaction phenomena.  Super-Kamiokande IV
has reported a $^8$B $\nu$ flux  measurement, ($2.34 \pm 0.03 \pm 0.04) \times
10^6$/cm$^2$s \cite{SKIV}, that continues  the progress toward high precision:
combining in quadrature the statistical  and systematic errors, one finds that
the Super-Kamiokande uncertainty is now  about 2\%.  The SNO combined analysis
of solar neutrino data from all phases, has significantly narrowed the allowed
range  for the  mixing angle  $\theta_{12}$  \cite{SNO}, and  new reactor  and
accelerator neutrino results have  fixed the contributions of the sub-dominant
mixing  angle $\theta_{13}$  \cite{theta13}.  A  new round  of  standard solar
model (SSM) calculations has  been completed to explore competing compositions
that  optimize SSM  agreement either  with the  solar interior  properties (as
determined  from helioseismic  mappings  of  the sound  speed)  or with  solar
surface properties (the interpretation of photoabsorption lines using our most
sophisticated  model  of the  Sun's  atmosphere)  \cite{PS2008}.  The  nuclear
physics  of the  SSM has  also been  updated, with  completion of  the nuclear
astrophysics community's  second decadal evaluation of SSM  S-factors and weak
interaction  rates  \cite{SFII}.  Finally,  Borexino  has  produced a  precise
(4.5\%)  measurement of  the 862  keV neutrinos  from $^7$Be  decay  and first
results  on  the  pep  flux,  following its  successful  calibration  campaign
\cite{Borexino}.  
The $^7$Be result has sharpened the ``luminosity constraint" on
the  pp/pep neutrinos, which currently provides our most precise
constraint on these fluxes if one assumes a steady-state Sun.

The SSM, despite  its relatively simple underlying physics,  depends on $\sim$
20 input parameters, including the  solar age and luminosity, the opacity, the
rate of diffusion,  the zero-age abundances of key elements (He,  C, N, O, Ne,
Mg, Si, S, Ar,  Fe), and the S-factors for the pp  chain (responsible for 99\%
of solar energy generation) and CN cycle.  While SSM neutrino flux predictions
generally change with variation in any  of these input parameters, it has been
recognized for some time that  flux predictions are strongly correlated with a
single output parameter, the  core temperature $T_c$ \cite{bahcall}.  That is,
a  multi-dimensional set  of variations  in  SSM input  parameters $\{  \Delta
\beta_j  \}$ from  the  SSM  best values  $\{  \beta_j^\mathrm{SSM} \}$  often
collapses to a one-dimensional dependence on $T_c$, where $T_c$ is an implicit
function of the variations $\{ \Delta \beta_j \}$: the effect of any variation
$\{ \Delta \beta_j \}$ on $\phi_i$ can be estimated simply from its effects on
$T_c$.   The dominance  of $T_c$  as  the controlling  parameter for  neutrino
fluxes  reflects  the  sensitivity  of Maxwellian-averaged  rates  $\langle  v
\sigma(E)  \rangle$  to  temperature.   Consequently, if  $\phi_{i_1}  \propto
T_c^{x_{i_1}}$ and  $\phi_{i_2} \propto T_c^{x_{i_2}}$, one  can form weighted
ratios  $\phi_{i_1}/\phi_{i_2}^{x_{i_1}/x_{i_2}}$ that are  nearly independent
of $T_c$ and thus nearly  independent of variations over the multi-dimensional
parameter space $\{ \Delta \beta_j \}$.
 
The situation becomes  more interesting when two fluxes,  say $\phi_{i_1}$ and
$\phi_{i_2}$,  in  addition to  their  common  dependence  on most  underlying
parameters $\{\beta_j\}$,  have a very  different dependence on  some specific
parameter, say  $\beta_1$.  In  that case, from  a weighted ratio  of observed
fluxes,  one might  be  able to  learn  something about  $\beta_1$, while  the
dependence on other  input parameters largely cancels out.   An example worked
out  previously  \cite{HS}  and   updated  below,  is  the  additional  linear
dependence of CN neutrino fluxes on  the primordial core number densities of C
and  N.   To the  extent  that SSM  uncertainties  exceed  the uncertainty  of
neutrino  flux measurements, new  information can  be extracted  from neutrino
measurements  --  in  this  case,  the  abundances of  C~and~N  in  the  Sun's
primordial  core.  Another  example we  will discuss  is the  possibility that
solar neutrino  fluxes can be  used to cross-check laboratory  measurements of
S-factors.
 
The  correlations between $\phi_i$  and $T_c$  are strong  but not  exact: for
example, as the neutrino-producing core is extended (with the extent depending
on  the neutrino  source),  fluxes must  depend  on an  integral  over a  core
temperature profile,  which cannot  be exactly proportional  to $T_c$  for all
variations.  Modern SSM calculations allow  one to address such issues, and to
include their effects  in the analysis.  Monte Carlo studies  can be done over
wide classes of  parameter variations $\{ \Delta \beta_j  \}$, determining not
only the best power-law descriptions of  the fluxes, but to also the extent of
reasonable variations around the power-law estimate.

The sensitivity  to parameter variations can  be be expressed in  terms of the
logarithmic partial derivatives $\alpha(i,j)$ evaluated for each neutrino flux
$\phi_i$ and each SSM input parameter $\beta_j$,
\begin{equation}
\alpha(i,j)  \equiv {\partial  \ln{\left[  \phi_i/\phi_i^\mathrm{SSM} \right]}
  \over \partial \ln{\left[ \beta_j / 
\beta_j^\mathrm{SSM}\right]}}
\end{equation}
where  $\phi_i^\mathrm{SSM}$ and  $\beta_j^\mathrm{SSM}$ denote  the  SSM best
values.  This  information, in combination with the  assigned uncertainties in
the  $\beta_j$,  then provides  an  estimate of  the  uncertainty  in the  SSM
prediction of $\phi_i$.  Here we employ the logarithmic partial derivatives of
\cite{PS2008},   which  were  evaluated   for  two   different  metallicities,
corresponding to the higher Z  composition of \cite{GS}, denoted GS98, and the
lower Z composition  of \cite{AGS}, denoted AGSS09. The  older GS98 abundances
were  obtained  from a  simple  analysis of  the  solar  atmosphere and  yield
excellent   agreement  with  interior   helioseismology.   The   newer  AGSS09
abundances,  obtained  from  a  more  sophisticated  3D  model  of  the  solar
atmosphere  that significantly  improves  the agreement  between measured  and
observed lines, are $\sim$ 30\% lower, and produce SSM sound speed profiles in
significant disagreement  with helioseismology.   The SSMs evolved  from these
compositions are denoted  SFII-GS98 and SFII-AGSS09 in this  paper, where SFII
(Solar  Fusion  II)  indicates  the   use  of  the  latest  nuclear  S-factors
\cite{SFII}.

The logarithmic partial derivatives for  the SFII-GS98 SSM are given in Tables
\ref{table:one} and  \ref{table:two}, divided as  in \cite{HS} into  two sets,
corresponding  to  nuclear   and  ``environmental"  $\beta_j$s.   The  nuclear
parameters are  the S-factors  for the  pp chain and  CN cycle:  S$_{11}$ (p+p
$\beta$      decay),     S$_{33}$      ($^3$He($^3$He,pp)$^4$He),     S$_{34}$
($^3$He($^4$He,$\gamma$)$^7$Be),  S$_{17}$ ($^7$Be(p,$\gamma$)$^8$B), S$_{e7}$
($^7$Be electron capture),  and S$_{114}$ ($^{14}$N(p,$\gamma$)$^{15}$O).  The
``environmental"  parameters  are  those  that directly  influence  the  local
temperature in the Sun, e.g., through their effects on evolution, the opacity,
or SSM boundary conditions.  They  include the luminosity $L_\odot$, the Sun's
age, the  diffusion parameter,  and the opacity.   They also include  the mass
fractions of the principal  solar metals, C, N, O, Ne, Mg,  Si, S, Ar, and Fe,
which have a significant influence on the opacity due to the strength of free
$\leftrightarrow$ bound transitions.

\begin{table*}
\caption{Partial derivatives $\alpha(i,j)$ of  neutrino fluxes with respect to
  solar  environmental  parameters  and  S-factors.   Table  entries  are  the
  logarithmic partial  derivatives $\alpha(i,j)$ of the  solar neutrino fluxes
  $\phi_i$  with respect  to the  indicated solar  model  parameter $\beta_j$,
  taken  about the  SFII-GS98  SSM best  values  \cite{PS2008}.  Several  flux
  ratios that reduce the solar environmental factors are shown.}
\label{table:one}
\begin{tabular}{|l|cccc|cccccc|}
\hline
 &   \multicolumn{4}{c|}  {~~~~~~~~~~~Environmental   $\beta_j$~~~~~~~~~~~}  &
\multicolumn{6}{c|}{~~~~~~~~~~~~~~~~~~~~Nuclear $\beta_j$~~~~~~~~~~~~~~~~~~~~}
\\ 
 \hline 
Source~~~~~~~~~~~~~~~~~~  &  ~$L_\odot$  ~&  Opacity &  ~~Age~~  &  Diffusion&
$S_{11}$ & $S_{33}$ & $S_{34}$ & $S_{17}$ & $S_{e7}$ & $S_{114}$ \\ 
\hline
$\phi$(pp) & 0.766 & -0.112 &-0.100 &  -0.013 & 0.105 & 0.034 & -0.067 & 0.000
& 0.000 & -0.007 \\
$\phi$(pep) &  0.989 & -0.318 &  -0.024 & -0.019 &  -0.217 & 0.049  & -0.097 &
0.000 & 0.000 & -0.010 \\
$\phi$($^7$Be) &  3.434 & 1.210 &  0.760 & 0.126 &  -1.024 & -0.428  & 0.853 &
0.000 & 0.000 & -0.001 \\
$\phi$($^8$B) &  6.914 & 2.611  & 1.345 &  0.267 & -2.651  & -0.405 &  0.806 &
1.000 & -1.000 & 0.007 \\
$\phi$($^{13}$N) & 4.535 &  1.487 & 0.932 & 0.337 & -2.166  & 0.031 & -0.062 &
0.000 & 0.000 & 0.747 \\
$\phi$($^{15}$O) & 5.942 &  2.034 & 1.364 & 0.382 & -2.912  & 0.024 & -0.052 &
0.000 & 0.000 & 1.000 \\
$\phi$($^7$Be)/$\phi$($^8$B)$^{0.465}$~~&  0.219 &  0.002 &  0.135 &  -0.004 &
0.209 & -0.240 & 0.478 & -0.465 & 0.465 & -0.004 \\ 
$\phi$($^{13}$N)/$\phi$($^8$B)$^{0.576}$~~ & 0.553 &  -0.017 & 0.157 & 0.183 &
-0.639 & 0.264 & -0.526 & -0.576 & 0.576 & 0.743 \\ 
$\phi$($^{15}$O)/$\phi$($^8$B)$^{0.785}$~~ & 0.515 &  -0.016 & 0.308 & 0.172 &
-0.831 & 0.342 & -0.685 & -0.785 & 0.785 & 0.995 \\ 
$\phi$($^{13}$N)/$\phi$($^{15}$O)$^{0.776}$~~  & -0.075  & -0.091  &  -0.126 &
0.041 & 0.093 & 0.012 & -0.022 & 0.000 & 0.000 & -0.029 \\ 
\hline \hline
\end{tabular}
\end{table*}

\begin{table*}
\caption{As  in  Table  \ref{table:one},   but  for  the  partial  derivatives
  $\alpha(i,j)$ with  respect to the  fractional abundances of  the primordial
  heavy elements.  .}
\label{table:two}
\begin{tabular}{|l|cc|ccccccc|}
\hline
 &       \multicolumn{2}{c|}      {~~~~~C,      N       $\beta_j$~~~~~}      &
\multicolumn{7}{c|}{~~~~~~~~~~~~~~~Environment                        Abundance
  $\beta_j$~~~~~~~~~~~~~~~} \\ 
 \hline
Source ~~~~~~~~~~~~~~~~~~ & C & N & O & Ne & Mg & Si & S & Ar & Fe \\
\hline
$\phi$(pp) & -0.008  & -0.002 & -0.006 &  -0.005 & -0.004 & -0.010  & -0.007 &
-0.002 & -0.021 \\ 
$\phi$(pep) & -0.016 &  -0.003 & -0.012 & -0.006 & -0.003  & -0.013 & -0.015 &
-0.005 & -0.062 \\ 
$\phi$($^7$Be) & 0.002 & 0.001 & 0.062 & 0.055 & 0.050 & 0.104 & 0.076 & 0.019
& 0.207 \\ 
$\phi$($^8$B) & 0.027 & 0.007 & 0.139 &  0.109 & 0.092 & 0.192 & 0.140 & 0.035
& 0.502 \\ 
$\phi$($^{13}$N) &  0.856 & 0.165 &  0.082 & 0.058 &  0.049 & 0.111  & 0.081 &
0.021 & 0.294 \\ 
$\phi$($^{15}$O) &  0.815 & 0.217 &  0.112 & 0.081 &  0.069 & 0.150  & 0.109 &
0.028 & 0.397 \\ 
$\phi$($^7$Be)/$\phi$($^8$B)$^{0.465}$~~& -0.011  & -0.002 & -0.003  & 0.004 &
0.007 & 0.015 & 0.011 & 0.003 & -0.026 \\ 
$\phi$($^{13}$N)/$\phi$($^8$B)$^{0.582}$~~ & 0.840 &  0.161 & 0.002 & -0.005 &
-0.004 & 0.000 & 0.000 & 0.001 & 0.005 \\ 
$\phi$($^{15}$O)/$\phi$($^8$B)$^{0.785}$~~ & 0.794 &  0.212 & 0.003 & -0.005 &
-0.003 & -0.001 & -0.001 & 0.001 & 0.003 \\ 
$\phi$($^{13}$N)/$\phi$($^{15}$O)$^{0.776}$~~  &  0.224   &  -0.003  &  -0.005
&-0.005 & -0.005 & -0.005 & -0.004 & -0.001 & -0.014 \\ 
\hline \hline
\end{tabular}
\end{table*}

The partial derivatives allow one to define the power-law dependencies of neutrino
fluxes, relative to the SSM best-value prediction $\phi_i^\mathrm{SSM}$
\begin{equation}
{   \phi_i  \over   \phi_i^\mathrm{SSM}}  =   \prod_{j=1}^N  x_j^{\alpha(i,j)}
\mathrm{~~where~~} x_j \equiv {\beta_j \over \beta_j^\mathrm{SSM}} 
\label{eq:prod}
\end{equation}
and  where  the  product  extends  over  19 SSM  input  parameters  of  Tables
\ref{table:one} and \ref{table:two}.  These derivatives determine how SSM flux
predictions will vary, relative to $\phi_i^\mathrm{SSM}$, as the $\beta_j$ are
varied from their SSM best values.

Our first  example of the use  of the logarithmic  partial derivatives follows
\cite{HS},  though the  analysis here  differs in  certain respects.   The Sun
produces about 1\% of its energy through the CN-I cycle, which produces $\nu$s
from    the    reactions    ${}^{13}\mathrm{N}(\beta^+)^{13}\mathrm{C}$    and
$^{15}\mathrm{O}(\beta^+)^{15}\mathrm{N}$,   with   respective   $\beta$-decay
endpoints of 1.20 MeV and 1.73 MeV.  Their fluxes in the SFII-GS98 SSM are
\begin{eqnarray}
\phi(^{13}\mathrm{N})&=&2.96 (1 \pm 0.14) \cdot 10^8/\mathrm{cm^2s} \nonumber \\
\phi(^{15}\mathrm{O})&=&2.23(1 \pm 0.15) \cdot 10^8/\mathrm{cm^2s}.
\label{eq:CNnus}
\end{eqnarray}
The primordial C and N in the  solar core are the catalysts for the conversion
of four protons to ${}^4$He via the CN cycle: the CN cycle alters the ratio of
C to  N as its burns  into equilibrium, but  does not change the  total number
density  of  C+N.   The  additional  linear  dependence  of  the  CN-cycle  on
metallicity, due to this dependence on  primordial C and N, can be isolated by
forming  a ratio of  fluxes that  is effectively  independent of  $T_c$, under
variations  in  all  other  SSM  parameters.  The  appropriate  ratio  can  be
identified  either   by  SSM  Monte   Carlo  studies,  at   considerable  cost
numerically, or  estimated from the  logarithmic partial derivatives.   It was
shown in \cite{HS} that the two approaches yield essentially the same answer.

The  neutrino  flux ratio  identified  in this  way  has  the requisite  metal
sensitivity to distinguish GS98 abundances from those of AGSS09, resolving the
solar  abundance problem.   More fundamentally,  it will  allow us  to  make a
important  test  of  a  key  SSM  assumption,  that  the  primordial  Sun  was
homogeneous when nuclear burning began -- an assumption not obviously correct,
given  what  we  have learned  in  the  past  decade about  large-scale  metal
segregation in the protoplanetary disk \cite{HS}, but nevertheless critical to
the SSM,  which uses solar  surface abundances to  fix core abundances  in the
primordial Sun.   Several groups have discussed relaxation  of this assumption
as  a possibility  for reconciling  helioseismic data  with  AGSS09 abundances
\cite{G06,CVR07,HS,GM10}.

The CN-cycle neutrino fluxes can be used as a direct probe of the core C~and~N
abundances only to the extent  that other SSM uncertainties can be controlled.
Uncertainties  in  S-factors  can  in  principle be  improved  through  better
laboratory measurements.  In  contrast, there may be no  effective strategy to
reduce   the  ``environmental"   uncertainties.   These   uncertainties  often
primarily affect the core temperature.  For example, when metal abundances are
varied, the SSM core temperature  responds to the resulting changes in opacity
and mean molecular weight: high metallicity cores are hotter.  Neutrino fluxes
also   respond,   reflecting  their   underlying   power-law  dependences   on
temperature.

The  dependence of the  fluxes on  environmental and  other parameters  can be
determined  from the  logarithmic  derivatives of  Tables \ref{table:one}  and
\ref{table:two},
\begin{align}
{\phi(^{13}\mathrm{N})\over \phi(^{13}\mathrm{N})^\mathrm{SSM}} =
\left[        L_\odot^{4.535}       O^{1.487}        A^{0.932}       D^{0.337}
  \right] \hspace{1.8cm}\nonumber \\ 
\times      \left[      \mathrm{S}_{11}^{-2.166}      ~\mathrm{S}_{33}^{0.031}
  ~\mathrm{S}_{34}^{-0.062} 
~\mathrm{S}_{17}^{0.0}~     \mathrm{S}_{e7}^{0.0}    ~\mathrm{S}_{114}^{0.747}
  \right] \hspace{1.7cm}\nonumber \\ 
\times      \left[      x_C^{0.856}      x_N^{0.165}      x_\mathrm{O}^{0.082}
  x_\mathrm{Ne}^{0.058} x_\mathrm{Mg}^{0.049} 
x_\mathrm{Si}^{0.111}        x_\mathrm{S}^{0.081}        x_\mathrm{Ar}^{0.021}
x_\mathrm{Fe}^{0.294} \right]~~~~~~~~ 
\label{eq:13N}
\end{align}
where    each    parameter   on    the    right-hand    side   represents    a
$\beta_j/\beta_j^\mathrm{SSM}$.  The  luminosity, opacity, solar  age, and the
diffusion parameters are denoted by $L_\odot$,  $O$, $A$, and $D$, while S and
$x$ denote S-factor or abundance ratios.  Similarly,
\begin{align}
{\phi(^{15}\mathrm{O})\over   \phi(^{15}\mathrm{O})^\mathrm{SSM}}   =   \left[
  L_\odot^{5.942}   O^{2.034}  A^{1.364}   D^{0.382}   \right]  \hspace{1.8cm}
\nonumber \\ 
 \times      \left[      \mathrm{S}_{11}^{-2.912}~     \mathrm{S}_{33}^{0.024}
   ~\mathrm{S}_{34}^{-0.052} ~\mathrm{S}_{17}^{0.0}~\mathrm{S}_{e7}^{0.0}~ 
\mathrm{S}_{114}^{1.00} \right] \hspace{1.7cm}  \nonumber \\
\times \left[x_C^{0.815} x_N^{0.217}x_\mathrm{O}^{0.112} x_\mathrm{Ne}^{0.081}
  x_\mathrm{Mg}^{0.069} 
x_\mathrm{Si}^{0.150}        x_\mathrm{S}^{0.109}        x_\mathrm{Ar}^{0.028}
x_\mathrm{Fe}^{0.397} \right].~~~~~~~~ 
\label{eq:15O}
\end{align}
The $^{15}$O $\nu$s are of  more interest experimentally, because their higher
energy  provides a  window for  observation  in a  scintillation detector,  as
discussed  in \cite{HS}.   From these  expressions and  from  the ``reasonable
ranges" for  input SSM  parameters given in  Table \ref{table:three},  one can
then  identify the  principal  sources  of SSM  uncertainty  in neutrino  flux
predictions.   The ranges  assigned in  Table \ref{table:three}  to  the metal
abundances are of  particular concern: the large differences  between GS98 and
AGSS09  reflect the  tension between  helioseismology and  3D modeling  of the
solar atmosphere.  For  each metal, we assign to  its abundance an uncertainty
formed by two contributions.   On one hand, following \cite{bahcallserenelli},
a {\it  systematic} component  based on the  differences between the  GS98 and
AGSS09 compositions given by 
\begin{equation}
{\Delta \beta_i \over \beta_i} = 2 \left| {\mathrm{Abundance}_i^{\mathrm{GS98}}
  - \mathrm{Abundance}_i^{\mathrm{AGSS09}} \over 
\mathrm{Abundance}_i^{\mathrm{GS98}}                                         +
  \mathrm{Abundance}_i^{\mathrm{AGSS09}}} \right|
\end{equation}
to which we add in quadrature the {\it observational} uncertainty taken as the
uncertainty quoted in the  latest solar abundance compilation \cite{AGS}. This
is a  conservative approach  for assigning abundance  uncertainties but  it is
appropriate for  the present work, as it  leads to robust upper  limits to the
precision with  which solar neutrino experiments can  constrain solar interior
properties.

\begin{table*}
\caption{Estimated 1$\sigma$ uncertainties  in solar (from Bahcall, Serenelli,
  and Basu \cite{BSB} and Fiorentini  and Ricci \cite{fr}) and nuclear physics
  (from  Adelberger  {\it  et   al.}  \cite{SFII})  uncertainties,  and  their
  influence  on flux  predictions, computed  from the  partial  derivatives of
  Table \ref{table:one}. }
\label{table:three}
\begin{tabular}{|l|c|ccccc|}
\hline
 & & & & & & \\[-.12in]
$\beta_j$~~~ &  Value & ~~${\Delta \beta_j \over  \beta_j}$(\%)~~ & ~~${\Delta
  \phi(^8\mathrm{B})   \over    \phi(^8\mathrm{B})}$   (\%)~~   &   ~~${\Delta
  \phi(^7\mathrm{Be}) \over \phi(^7\mathrm{Be})}$(\%)~~ & 
~~${\Delta   \phi(^{13}\mathrm{N})   \over   \phi(^{13}\mathrm{N})}$(\%)~~   &
~~${\Delta \phi(^{15}\mathrm{O}) \over \phi(^{15}\mathrm{O})}$ (\%)~~\\[.05in] 
\hline
L$_\odot$ & 3.842$\times$10$^{33}$ ergs/s & 0.4 & 2.8 & 1.4 & 1.8 & 2.4   \\ 
Opacity  & 1.0 & 2.5 & 6.5 & 3.0 & 3.7 & 5.1 \\
Age  & 4.57 Gyr & 0.44 &  0.59 & 0.33 & 0.41 & 0.60 \\
Diffusion  & 1.0 & 15.0 &  4.0 & 1.9 & 5.1 & 5.7 \\
p+p & (4.01$\pm0.04)\times$10$^{-25}$ MeV b & 1.0 & 2.6 & 1.0 & 2.2 & 2.9 \\
 $^3$He+$^3$He & (5.21$\pm$0.27) MeV b & 5.2 & 2.1 & 2.2 & 0.16 & 0.12 \\
$^3$He+$^4$He & (0.56$\pm$0.03) MeV b & 5.4 &  4.3 & 4.6 & 0.33 & 0.28 \\
 p+$^7$Be & (20.8$\pm$1.6) eV b & 7.7 & 7.7 & 0.0 & 0.0 & 0.0 \\
e+$^7$Be &  & 2.0  & 2.0 & 0.0 & 0.0 & 0.0 \\
p+$^{14}$N & (1.66$\pm$0.12) keV b & 7.5 &  0.05 & 0.0 & 5.6 & 7.5 \\
\hline \hline
\end{tabular}
\end{table*}

\begin{table*}
\caption{Estimated  1$\sigma$  historical  (``conservative") uncertainties  in
  AGSS98    abundances,    as     defined    in    Bahcall    and    Serenelli
  \cite{bahcallserenelli}.   The corresponding  uncertainties in  the neutrino
  fluxes are computed from the partial derivatives of Table \ref{table:two}.}
\label{table:four}
\begin{tabular}{|l|ccccc|}
\hline
 & & & & & \\[-.12in]
$\beta_j$~~~&   ~~${\Delta   \beta_j   \over  \beta_j}$(\%)~~   &   ~~${\Delta
  \phi(^8\mathrm{B})    \over    \phi(^8\mathrm{B})}$(\%)~~    &    ~~${\Delta
  \phi(^7\mathrm{Be}) \over \phi(^7\mathrm{Be})}$(\%)~~ & 
~~${\Delta   \phi(^{13}\mathrm{N})   \over   \phi(^{13}\mathrm{N})}$(\%)~~   &
~~${\Delta \phi(^{15}\mathrm{O}) \over \phi(^{15}\mathrm{O})}$(\%)~~ \\[.05in] 
\hline
 C & 24.6 & 0.66 & 0.05 & 21.1 & 20.1 \\ 
 N & 24.6 & 0.17 & 0.02 & 4.1 & 5.3\\
 O & 35.0 & 4.9 & 2.2 & 2.9 & 3.9 \\
 Ne & 45.3 & 4.9 & 2.5  & 2.6 & 3.7 \\
 Mg & 11.8 & 1.1 & 0.59 & 0.58 & 0.81 \\
  Si & 11.8 & 2.3 & 1.2 & 1.3 & 1.8 \\
  S & 13.8 & 1.9 & 1.0 & 1.1 & 1.5\\
 Ar & 34.9 & 1.2 & 0.66 & 0.73 & 0.98\\
  Fe & 11.8 & 5.9 & 2.4 & 3.5 & 4.7\\
\hline \hline
\end{tabular}
\end{table*}

It is reasonable to treat the effects of abundances in Eqs. (\ref{eq:13N}) and
(\ref{eq:15O})  as  an overall  scaling  of  metallicity,  as the  differences
between the GS98  and AGSS09 abundances are effectively  systematic in the net
metallicity.   With   this  assumption,   the  dominant  SSM   uncertainty  in
Eq. (\ref{eq:15O}) is the core  abundance of \hbox{C + N} which, if
changed systematically over a  range equivalent to the GS98-AGSS09 difference,
alters the $^{15}$O neutrino flux by  30.7\%.  This is the sensitivity we want
to exploit,  in using CN neutrinos as  a probe of core  metallicity.  The next
largest certainty comes from the 11 environmental parameters, 16.5\%: thus the
environmental  uncertainties are  the  primary factor  inhibiting  our use  of
neutrinos  as  a  probe  of  composition.  The  uncertainty  coming  from  the
S-factors, 7.7\%, is entirely  dominated by S$_{114}$, which alone contributes
7.2\%.

Now  the  nuclear  physics  uncertainties  can  be  reduced  with  effort:  in
\cite{SFII} possible  steps to improve existing measurements  of S$_{114}$ are
described.  But we have less  control over the environmental parameters, so an
alternative strategy  is needed  to address these  uncertainties.  We  use the
well-measured flux of  $^8$B as a solar thermometer, to remove  as much of the
environmental dependence as possible.

We  form a weighted  ratio of  the $^{15}$O  $\nu$ and  $^8$B $\nu$  fluxes to
eliminate the dependence on $T_c$ to the extent possible, and thus to minimize
the  dependence   on  10  of   the  11  environmental  parameters   of  Tables
\ref{table:one}  and   \ref{table:two}:  we  do  not   include  the  diffusion
coefficient,  as this  parameter  plays  a special  role  in the  relationship
between contemporary  flux measurements and the primordial  abundances we seek
to  constrain.  The CN  neutrino fluxes  are more  sensitive to  diffusion, as
Table \ref{table:one} shows.  All neutrino fluxes respond similarly to changes
in core temperature induced  by gravitational settling.  However, the $^{15}$O
flux has an additional dependence  on changes in the $^{12}$C and $^{14}$N
core abundances, as the rate is proportional to those abundances, for constant
temperature.  Thus the analysis is done in a way that isolates this additional
dependence.

To exploit the well-measured flux of  $^8$B neutrinos as a thermometer in this
way, one must determine the linear correlations between ln($\phi(^{13}$N)) and
ln($\phi(^8$B))  and between  ln($\phi(^{15}$O))  and ln($\phi(^8$B)).   While
this  can be  done  by direct  Monte  Carlo SSM  calculations (see  discussion
below), it was shown in \cite{HS}  that such an exercise is largely equivalent
to minimizing the  dependence on net logarithmic derivatives.  The solution to
the minimization is a power  law of the N observables, $\prod_{i=1}^N({ \phi_i
  \over  \phi_i^\mathrm{SSM}})^{b^{k}_{i}}$, with exponents  $b^{k}_{i}$ given
by the  eigenvector with minimum  eigenvalue of the nuisance  parameters error
matrix \cite{stat}
\begin{equation}
{\cal   M}_{il}   =  \sum_{j=1}^{n}\bigg(\frac{\Delta\beta_j}{\beta_j}\bigg)^2
\alpha(i,j) \alpha(l,j)\,. 
 \label{eq:eigenvalue}
\end{equation}
The  computation of  the matrices  ${\cal M}_{^{8}\mathrm{B},^{13}\mathrm{N}}$
and   ${\cal  M}_{^{8}\mathrm{B},^{15}\mathrm{O}}$   is   
straightforward.   
The direction of the smallest eigenvalue is, respectively,
\begin{align}
 {\phi(^{13}\mathrm{N})\over  \phi(^{13}\mathrm{N})^\mathrm{SSM}}\Big/  \left[
   {\phi(^{8}\mathrm{B})     \over     \phi^\mathrm{SSM}(^{8}\mathrm{B})     }
   \right]^{0.576} = x_C^{0.840} x_N^{0.161} D^{0.183} 
\hspace{1.5cm} \nonumber \\ \times \left[ L_\odot^{0.553} O^{-0.017} A^{0.157}
  \right] \hspace{2.3cm}  \nonumber \\ \times  \left[ \mathrm{S}_{11}^{-0.639}
  ~\mathrm{S}_{33}^{0.264}                            ~\mathrm{S}_{34}^{-0.526}
  ~\mathrm{S}_{17}^{-0.576}~ \mathrm{S}_{e7}^{0.576} ~\mathrm{S}_{114}^{0.743}
  \right]   \hspace{1.7cm}\nonumber  \\  \times   \left[  x_\mathrm{O}^{0.002}
  x_\mathrm{Ne}^{-0.005}     x_\mathrm{Mg}^{-0.004}     x_\mathrm{Si}^{0.0}
  x_\mathrm{S}^{0.0}        x_\mathrm{Ar}^{0.001}        x_\mathrm{Fe}^{0.005}
  \right] \hspace{1.4cm}
\end{align}
and
\begin{align}
 {\phi(^{15}\mathrm{O})\over  \phi(^{15}\mathrm{O})^\mathrm{SSM}} \Big/ \left[
   {\phi(^{8}\mathrm{B})     \over     \phi^\mathrm{SSM}(^{8}\mathrm{B})     }
   \right]^{0.785}= x_C^{0.794} x_N^{0.212} D^{0.172} 
 \hspace{1.5cm}   \nonumber  \\   \times  \left[   L_\odot^{0.515}  O^{-0.016}
   A^{0.308}    \right]   \hspace{2.3cm}    \nonumber    \\   \times    \left[
   \mathrm{S}_{11}^{-0.831} ~\mathrm{S}_{33}^{0.342} ~\mathrm{S}_{34}^{-0.685}
   ~\mathrm{S}_{17}^{-0.785}~                           \mathrm{S}_{e7}^{0.785}
   ~\mathrm{S}_{114}^{0.995} \right]  \hspace{1.7cm}\nonumber \\ \times \left[
   x_\mathrm{O}^{0.003}      x_\mathrm{Ne}^{-0.005}     x_\mathrm{Mg}^{-0.003}
   x_\mathrm{Si}^{-0.001}       x_\mathrm{S}^{-0.001}      x_\mathrm{Ar}^{0.001}
   x_\mathrm{Fe}^{0.003} \right] \hspace{1.4cm}
\label{eq:ratio}
\end{align}

The important dependences on opacity and metallicity (other than C and N) have
been almost  entirely removed.  While  some residual dependence  on luminosity
and age  remains, these parameters have relatively  small uncertainties.  Once
the $^8$B neutrino thermometer has  removed the environmental effects, we find
that the $^{15}$O neutrino flux  varies linearly under scaling of the $^{12}$C
and $^{14}$N abundances ($0.794 + 0.212 = 1.006 \sim 1$).  This dependence can
be made more explicit in Eq. (\ref{eq:ratio}) by the replacement
\begin{equation}
x_C^{0.794} x_N^{0.212} \Rightarrow \left[ { N_\mathrm{C} + N_\mathrm{N} \over
    N_\mathrm{C}^\mathrm{SSM} + N_\mathrm{N}^\mathrm{SSM}} \right]
\end{equation}
where  $N_\mathrm{C}$ and  $N_\mathrm{N}$ are  the number  densities of  C and
N. That is, the $^{15}$O neutrino  flux depends effectively only on the sum of
the number densities.  The exponents appearing in Eq.  (\ref{eq:ratio}) depend
weakly on the  SSM about which the variations are made.   We use the SFII-GS98
SSM     where     $N_\mathrm{C}^\mathrm{SSM}/N_\mathrm{N}^\mathrm{SSM}    \sim
0.80/0.20$; the same ratio occurs  in solar models using the solar composition
from \cite{AGS}.

For $^{15}$O, the case of  most interest experimentally, the observable on the
left hand  side responds  linearly to any  scaling of  the N and  C primordial
abundances.   Diffusion,  using   Table  \ref{table:three},  creates  a  2.6\%
uncertainty  in  relating contemporary  flux  measurements  to the  primordial
abundance.  This  2.6\% is virtually all  that remains of  the original 16.5\%
SSM  environmental  uncertainty  of  Eq. (\ref{eq:15O}):  the  $^8$B  neutrino
thermometer  has reduced the  uncertainties associated  with the  remaining 10
parameters to below  0.35\%.  The third term on  the right, contributions from
the S-factors, is now the dominant theoretical uncertainty in the relationship
between primordial C+N and  neutrino flux measurements, contributing 10.6\% to
the error budget.

\begin{figure*}
\begin{center}
\includegraphics[width=14cm]{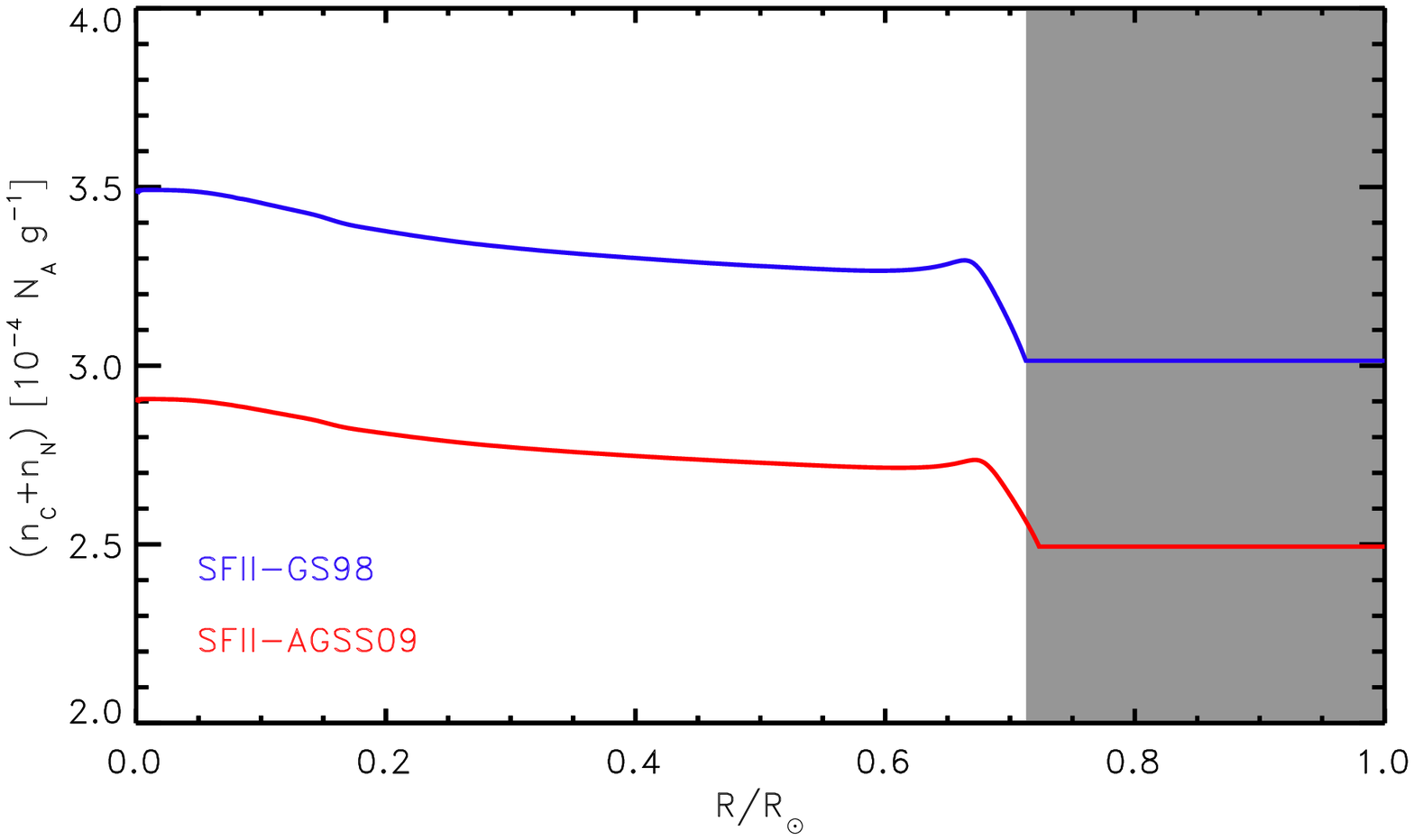}
\caption{The  modern  Sun's  carbon  plus  nitrogen  number  profiles  in  the
  SFII-GS98 and  SFII-AGSS09 SSMs, showing  the effects of diffusion  over 4.6
  Gyr of stellar evolution. The shaded area denotes the convective envelope.}
\label{fig:metals}
\end{center}
\end{figure*}

The explicit  treatment of diffusion, effectively grouping  diffusion with the
C+N abundance,  differs from the original  work of \cite{HS}.   This choice is
made for simple  physical reasons, that neutrino flux  measurements respond to
contemporary  core abundances,  yet the  parameters needed  in the  SSM, which
describes  the  Sun's  evolution  from  the  onset  of  nuclear  burning,  are
primordial.  Thus  the relationship we  establish between primordial  core C+N
and contemporary CN neutrino fluxes  has a dependence on diffusion that should
be made explicit, as we have  done here.  Indeed, the effects of diffusion are
not  inconsequential: the  SFII-GS98  and SFII-AGSS09  SSM  metal profiles  of
Fig.  \ref{fig:metals} show  that diffusion  over 4.6  Gyr of  solar evolution
leads  to   nontrivial  structures.    Fortunately  for  our   present  goals,
helioseismology is sensitive  to He and metal diffusion:  the 15\% uncertainty
on the diffusion coefficient (see Table  III) is a credible limit on diffusion
uncertainties because of helioseismic constraints.

\begin{figure*}
\begin{center}
\includegraphics[width=14cm]{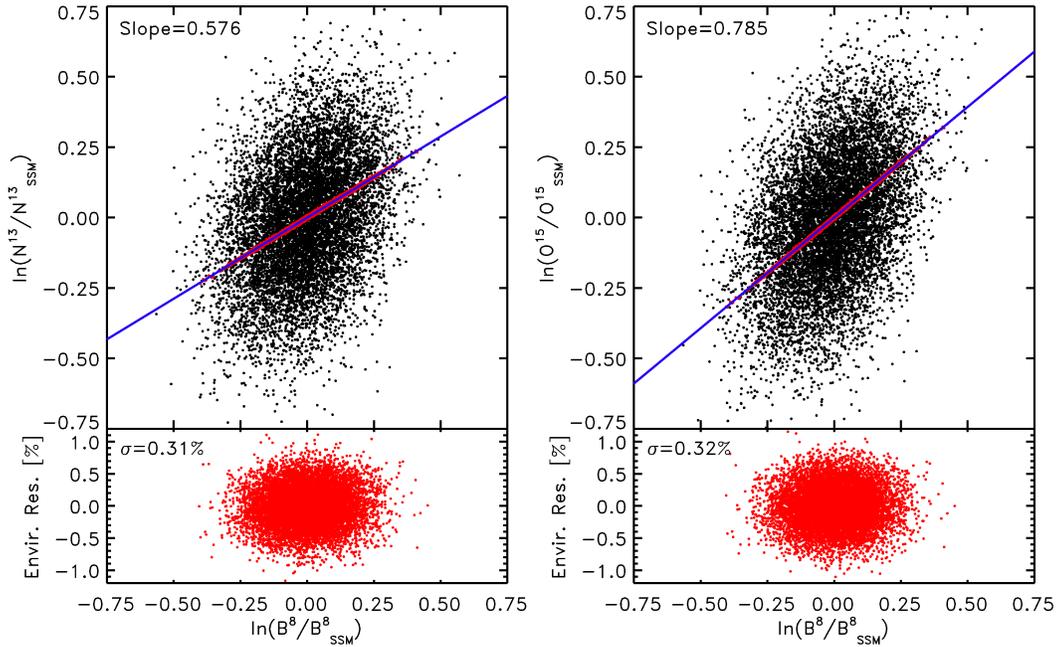}
\caption{Solar neutrino fluxes from models in which all the parameters (black)
  or 10 of the 11  environmental parameters (red) -- the diffusion coefficient
  is  held  fixed   --  are  varied.   At  this   resolution  red  points  are
  indistinguishable from  a line.  The  two upper panels show  the correlation
  between the $^8$B flux and the  two CN-cycle neutrino fluxes.  The slopes of
  the correlations  between fluxes when  only 10 environmental  parameters are
  varied are  given in  the plots.   The residuals from  the fits,  0.31\% and
  0.32\%, are shown in the lower panels.  The results can be compared to those
  of Fig.   3 in \cite{HS}, where  residuals of 2.8\% and  2.6\% were obtained
  for  the $^{13}$N  and  $^{15}$O fluxes,  respectively,  when diffusion  was
  included as an 11th environmental parameter.}
\label{fig:correlation}
\end{center}
\end{figure*}

Once this dependence  on diffusion is separated out,  it becomes apparent that
almost  all of  the  residual SSM  ``environmental"  dependence identified  in
\cite{HS} --  variations in 11 SSM  parameters producing a  net uncertainty of
2.6\% -- is due to the diffusion coefficient.  The correlations illustrated in
Fig. 3  of Ref. \cite{HS} are  redone in Fig.  \ref{fig:correlation}, with the
removal  of this  one  parameter.  The  net  uncertainty due  to
uncorrelated  variations in  the remaining  10  parameters is  now reduced  to
0.3\%, as mentioned above.

The Super-Kamiokande measurement of the  $^8$B flux has reached a precision of
2\%.   Borexino has set  the strongest  constraint on  the CNO  solar neutrino
interaction  rate ($<$7.9  counts/(day x  100 ton)  at 95  \% C.L.)  and their
latest purification  campaign has resulted  in a much lower  background level,
what  opens  the   possibility  of  the  first  detection   of  CNO  neutrinos
\cite{BorexinoII}.  SNO+ has the potential  to measure the $^{15}$O flux to an
accuracy of about 10\% in three years of running, if the detector design goals
are reached \cite{SNO+}.  Thus the current theoretical $\sim 30$\% uncertainty
in  the  core C+N  abundance  could be  substantially  reduced  by a  neutrino
measurement.   In fact,  the limiting  uncertainty appears  to be  the nuclear
physics, specifically  S$_{17}$ (7.7\%) and S$_{114}$ (7.5\%).   Both of these
reactions  were  recently  evaluated  by the  nuclear  astrophysics  community
\cite{SFII}.  The uncertainty  in S$_{17}$ is dominated by  the theory used to
fit  and extrapolate  measurements  -- the  experimental  contribution to  the
S$_{17}$ error is  3.4\% \cite{SFII}.  As {\it ab initio}  methods may soon be
available  for  such  systems  \cite{marcucci}, the  situation  could  improve
substantially.  In the case of S$_{114}$ a program of needed work was outlined
in \cite{SFII}, including new measurements to constrain the transitions to the
6.79 and 6.17 MeV states in  $^{15}$O.  We conclude that it should be possible
to significantly reduce the overall uncertainty from the nuclear physics.

The expressions  above are valid  for the neutrino  fluxes at the  source.  We
need to account for the effects  of neutrino flavor conversion, as this alters
the  ratio of  detected  $^{15}$O to  $^8$B  neutrinos in  detectors based  on
$\nu_x$-e  scattering.  $^8$B neutrino  oscillation probabilities  are smaller
because their  energies correspond to  the matter dominated  flavor conversion
while the CN neutrinos are in  the vacuum oscillation regime with small matter
effects.    We  use   the  most   up-to-date  neutrino   oscillation  analysis
\cite{GonzalezGarcia}  to  estimate  the   uncertainty  due  to  the  neutrino
parameters, with  lowers the weak-interactions uncertainty in  our analysis to
$\pm$ 3\%.

The  analysis suggests  that  a neutrino  determination  of the  \hbox{C +  N}
content of  the core at  a confidence level  of $\sim$ 10\% is  quite feasible
with  the   future  measurements.   This  assumes  a   7\%  $^{15}$O  neutrino
measurement and  modest progress in lowering nuclear  physics uncertainties to
the  same  level.   This  should   be  compared  to  the  current  metallicity
controversy, $\sim$ 30\%.  Such a  measurement would also constitute the first
direct experimental test  of an important SSM assumption,  that the primordial
core and modern solar atmosphere  metallicities are the same, once corrections
are made  for the effects of  diffusion.  Other sources of  uncertainty -- the
Super-Kamiokande measurement of the $^8$B neutrino rate for elastic scattering
(ES), the SNO combined  analysis constraining weak interaction parameters, and
the influence of  diffusion (the one effect intrinsic to the  SSM that can not
be adequately subtracted  using the $^8$B neutrino thermometer)  -- are all of
minor importance, contributing to the error budget at $\sim$ 3\%.
 
There is a  second constraint on core composition that  could be obtained with
CN  neutrinos and  that is  inherently interesting  because the  observable is
exceptionally free of SSM uncertainties.  This constraint, however, requires a
measurement of the $^{13}$N neutrinos.  If the pep shoulder is seen and if the
level  of background  in the  ES  measurements can  be kept  low (or  reliably
subtracted),  the  remaining  counts  in  the $\sim$  1~MeV  region  could  be
associated with the $^{13}$N and $^{15}$O neutrinos. One important observation
is that  the relative contributions of  the two neutrino sources  to the total
rate  vary with  the electron  recoil energy,  due to  the  different neutrino
energy distributions.  This could allow  the experimenters to separate the two
CN-neutrino  flux  components.   In  Table  \ref{table:CNO_es},  we  show  the
relative contribution  to the  scattering rate as  a function of  the measured
recoil electron  energy interval.  We have  used the energy  resolution of the
Borexino  detector  and  the  $^{13}$N  and $^{15}$O  energy  distribution  of
neutrinos  given by  the SFII-GS98  model.   In the  interesting energy  range
between the  $^7$Be and pep shoulders,  the relative contribution  to the rate
varies by  one order of magnitude.  The  higher energy bins in  this range are
strongly dominated  by the $^{15}$O  neutrino flux contribution and  the error
estimate of  the flux  will be  comparable to the  experimental error  in this
energy region. The lower energy  bins have a significant contribution from the
$^{13}$N neutrino  flux.  Thus  this flux component  could also  be determined
from the  data, though with a larger  uncertainty due to the  need to separate
this component  from the dominant $^{15}$O contribution.   An important caveat
is  the assumption  there are  no unidentified  background sources  that might
mimic the $^{13}$N neutrino signal.

\begin{table*}
\caption{Ratio of the  scattering rate by $^{15}$O and  $^{13}$N neutrino with
  electrons in a Borexino-like detector.}
\label{table:CNO_es}
\begin{tabular}{lccccccccl}
\hline  &&&&&&& \\[-.12in]  E (MeV)  &~~ [0.70,0.75]~~  &~~  [0.75,0.80]~~ &~~
       [0.80,0.85] ~~&  ~~ [0.85,0.90]  ~~& ~~ [0.90,0.95]  ~~& ~~[0.95,1.0]~~
       \\[.05in]            $R(^{15}\mathrm{O})/           R(^{13}\mathrm{N})$
       &2.1&2.6&3.6&5.5&10.2&24.7&\\[.05in] \hline \hline
\end{tabular}
\end{table*}

As noted previously, the CN cycle has not reached equilibrium in the Sun apart
from its  central core.  The lifetime  of $^{14}$N, determined by  the rate of
$^{14}$N(p,$\gamma$), is less  than the solar age only  for $T_7 \gtrsim 1.33$
($T_7$ is the  temperature in $10^7$K).  But at  this temperature the lifetime
of $^{12}$C is  $\sim 2 \cdot 10^7$ years.  Thus  somewhat outside the central
core, say at $T_7 \sim 1.15$,  there will be very little $^{14}$N burning, and
also very  little $^{12}$C burning, as  the primordial carbon  would have been
consumed long ago.  Still further  outward, where $T \sim 10^7$K, the $^{12}$C
lifetime  is  comparable  to  the  solar  age.  This  is  the  region  in  the
contemporary Sun where primordial $^{12}$C  is being burned.  We conclude that
CN  neutrinos are  coming  from two  distinct  regions.  The  CN  cycle is  in
equilibrium  deep  in  the  core,  producing approximately  equal  numbers  of
$^{15}$O  and $^{13}$N neutrinos,  while well  away from  this region,  in the
cooler  outer  core at  $T  \sim 10^7$K,  primordial  $^{12}$C  is burning  to
$^{14}$N, producing only low-energy $^{13}$N neutrinos.

That  is,  the  unequal  fluxes  of  $^{13}$N and  $^{15}$O  neutrinos  are  a
reflection of  the burning of primordial  $^{12}$C in the outer  core.  We can
test for this effect by comparing these fluxes, treating the $^{15}$O neutrino
flux  as the  thermometer.  In  the exercise  to find  the  linear correlation
between  the   logarithmic  fluxes,  we   now  include  diffusion   among  the
environmental parameters,  as it should affect $^{13}$N  and $^{15}$O neutrino
rates almost equally.  We find
\begin{align}
 {\phi(^{13}\mathrm{N})\over  \phi(^{13}\mathrm{N})^\mathrm{SSM}}\Big/  \left[
   {\phi(^{15}\mathrm{O})     \over    \phi^\mathrm{SSM}(^{15}\mathrm{O})    }
   \right]^{0.776} = 
x_C^{0.224} x_N^{-0.003} \mathrm{S}_{112}^{-0.008} 
\hspace{2.1cm} 
\nonumber \\
\times     \left[    L_\odot^{-0.075}    O^{-0.091}     A^{-0.126}    D^{-0.041}
  \right] \hspace{5.2cm} \nonumber \\ 
\times      \left[      \mathrm{S}_{11}^{0.093}      ~\mathrm{S}_{33}^{0.012}
  ~\mathrm{S}_{34}^{-0.022} 
~\mathrm{S}_{17}^{0.0}~     \mathrm{S}_{e7}^{0.0}    ~\mathrm{S}_{114}^{-0.029}
  \right] \hspace{4.2cm}\nonumber \\ 
\times \left[  x_\mathrm{O}^{-0.005} x_\mathrm{Ne}^{-0.005} x_\mathrm{Mg}^{-0.005}
x_\mathrm{Si}^{-0.005}        x_\mathrm{S}^{-0.004}        x_\mathrm{Ar}^{-0.001}
x_\mathrm{Fe}^{-0.014} \right]. \hspace{2cm} \label{eq:n13o15}
\end{align}
The residual environmental uncertainty is  only $\sim$ 0.7\%.  For the nuclear
part we  have made explicit the very  small dependence on the  S-factor of the
$^{12}$C(p,$\gamma$)$^{13}$N reaction rate, ${\rm S_{112}}$. The total nuclear
uncertainty in the above expression is only 0.3\%. We note here the dependence
of this ratio on the nitrogen abundance is partially accidental.  The exponent
that minimizes  the environmental uncertainties,  0.776, is very close  to the
ratio  of the  partial  derivatives of  these  fluxes with  respect  to the  N
abundance 0.165/0.217=0.760  (see Table II).   However, we find that  even for
unrealistically large  variations of more  than a factor  of 2 in  the assumed
composition  of the  Sun, this  ratio varies  little, between  0.71  and 0.81.
Therefore, the  cancellation of the nitrogen  abundance in Eq.~\ref{eq:n13o15}
will always  occur at a level better  than 0.5\%. A similar  conclusion can be
drawn with respect to the  ${\rm S_{114}}$ astrophysical factor, which is also
accidentally cancelled for the same reason.

{\it Other constraints:} While we have focused on metallicity and the CN-cycle
neutrinos,  due to  the  troubling solar  abundance  problem, the  use of  SSM
power-law  temperature  dependences  to  extract parameter  constraints  is  a
general strategy  for exploiting  the Sun as  a laboratory.  For  example, the
primordial  $^4$He abundance  \cite{SB}  was recently  constrained using  very
similar  arguments.  Another  example we  discuss here  is the  possibility of
using the  SSM to cross-check laboratory measurements  of S-factors.  S$_{17}$
is an  important example because of  the relatively large  uncertainty in this
S-factor and because  of its importance to the branching  between the ppII and
ppIII cycles.  The  analysis is quite simple, a comparison  the the $^7$Be and
$^8$B  fluxes,  neither  of  which  has  any  anomalous  dependence  on  metal
abundances  or  diffusion.  Thus  we  can  optimize  over all  13  non-nuclear
parameters, with the anticipation that the  residuals will be small in each of
these.  Following the previous calculation, we find
\begin{align}
 {\phi(^{7}\mathrm{Be})\over  \phi(^{7}\mathrm{Be})^\mathrm{SSM}}\Big/  \left[
   {\phi(^{8}\mathrm{B})     \over     \phi^\mathrm{SSM}(^{8}\mathrm{B})     }
   \right]^{0.465} = 
 \hspace{5.0cm} \nonumber \\
 \left[    L_\odot^{0.219}     O^{-0.004}    A^{0.135}    D^{0.002}    \right]
 \times \hspace{4.0cm} \nonumber \\ 
 \left[ \mathrm{S}_{11}^{0.209} ~\mathrm{S}_{33}^{-0.240} ~\mathrm{S}_{34}^{0.479}
~\mathrm{S}_{17}^{-0.465}~  \mathrm{S}_{e7}^{0.465} ~\mathrm{S}_{114}^{-0.004}
   \right] \times \hspace{3.1cm}\nonumber \\ 
 \left[   x_\mathrm{C}^{-0.011}   x_\mathrm{N}^{-0.002}  x_\mathrm{O}^{-0.003}
   x_\mathrm{Ne}^{0.004} x_\mathrm{Mg}^{0.007} 
x_\mathrm{Si}^{0.015}        x_\mathrm{S}^{0.011}        x_\mathrm{Ar}^{0.003}
x_\mathrm{Fe}^{-0.026} \right] \nonumber \hspace{2.7cm} \\ 
\sim  \left[  {\mathrm{S}_{11}  \over \mathrm{S}_{33}}  \right]^{0.24}  \left[
  {\mathrm{S}_{34} \mathrm{S}_{e7} \over 
\mathrm{S}_{17}} \right]^{0.48} F_\mathrm{SSM}^\mathrm{nonnuclear} \hspace{4.0cm}
\end{align}

The  error  introduced by  grouping  the  astrophysical  factors in  the  last
expression    is    only   0.1\%.     In    this    expression   the    factor
$F_\mathrm{SSM}^\mathrm{nonnuclear}$ represents the  contributions from the 13
non-nuclear  uncertainties in  the  SSM:  using the  exponents  above and  the
$\beta_j$ of Tables III and IV, one finds that this contribution deviates from
unity  by $\sim  \pm$0.5\%, and  therefore  plays no  significant role.   Thus
effectively we  have a direct relationship between  neutrino flux measurements
and nuclear cross sections.  The left-hand side of Eq.  (11) is the product of
two    factors.     The    first,   $\left[    \mathrm{S}_{11}/\mathrm{S}_{33}
  \right]^{0.24}$,  is uncertain  to  1.3\%, using  the  evaluations of  Solar
Fusion II, with the error dominated by that in $\mathrm{S}_{33}$.  The second,
$  \left[ {\mathrm{S}_{34} \mathrm{S}_{e7}/  \mathrm{S}_{17}} \right]^{0.48}$,
is uncertain to  4.6\%, treating all uncertainties as  uncorrelated.  Thus the
left-hand  side of  Eq.  (11)  is $1  \pm 0.048$,  when all  uncertainties are
combined in quadrature.

The right-hand side can be evaluated from the results of global solar neutrino
flux analyses  that incorporate the neutrino oscillation  results important to
mixing angle determinations (as  the fluxes are the unoscillated instantaneous
ones)\cite{PS2008}.   The analysis  is  done in  terms  of the  normalizations
provided by  SFII-GS98 SSM best  values, for consistency with  the logarithmic
derivatives  we  employ:  $\phi(^7\mathrm{Be})=5.00 \times  10^9$/cm$^2$s  and
$\phi(^8\mathrm{B})=5.58  \times 10^6$/cm$^2$s.   The experimental  fluxes are
$4.82(1   \pm   0.045)  \times   10^9$/cm$^2$s   and  $5.00(1\pm0.03)   \times
10^6$/cm$^2$s.  Consequently  the left-hand side is $1.016(1  \pm 0.047)$.  If
the SFII-AGSS09 SSM best values are  used to normalize the left hand side, the
result is virtually unchanged, $1.015(1 \pm 0.047)$.

This result is  significant: the constraint imposed on  the ratio of S-factors
$\left[ \mathrm{S}_{11}/\mathrm{S}_{33} \right]^{0.24} \left[ {\mathrm{S}_{34}
    \mathrm{S}_{e7}/  \mathrm{S}_{17}} \right]^{0.48}$  relative to  SFII best
values, using all information  available from laboratory astrophysics, has the
same  precision  as  the  similar  ratio  we can  deduce  from  neutrino  flux
measurements, if we employ the SSM  to predict the dependence of the fluxes on
input S-factors, and  if we constrain all non-nuclear  parameters to vary only
within  the ranges allowed  by their  currently assigned  uncertainties.  This
result was achieved by identifying a specific ratio of $^7$Be and $^8$B fluxes
that the  SSM predicts will exhibit  the minimum uncertainty  to variations in
the 13 non-nuclear  parameters.  The two independent constraints  -- the left-
and right-hand sides of Eq. (11)  -- are in excellent agreement, a result that
reflects  the concordance  between neutrino  flux observations  and laboratory
nuclear  cross  section measurements,  in  the context  of  the  SSM.  As  the
uncertainty of the  neutrino-flux result for this S-factor  ratio is dominated
almost entirely by that for  the $^7$Be neutrino flux, further improvements in
the Borexino result (or new  results from a next-generation experiment such as
SNO+) would make the neutrino-flux S-factor constraint the more precise one.

While  this  test  of  concordance  between  neutrino  flux  measurements  and
laboratory  measurements of  S-factors  is our  main  point, one  can be  more
aggressive and  ask whether new  S-factor information can be  derived directly
from neutrino  flux measurements.  As the  most uncertain of  the S-factors is
S$_{17}$, what level  of precision is needed in  neutrino flux measurements to
improve our knowledge of this cross section?  Equation (11) can be rewritten
\begin{eqnarray}
\mathrm{S}_{17}  &=& \mathrm{S}_{34}  \mathrm{S}_{e7}  \left[ {\mathrm{S}_{11}
    \over            \mathrm{S}_{33}}           \right]^{0.5}           \left[
  F_\mathrm{SSM}^\mathrm{nonnuclear}\right]^{2.08} \nonumber \\ 
&\times& {\phi(^{8}\mathrm{B}) \over \phi^\mathrm{SSM}(^{8}\mathrm{B}) } \Big/
\left[     {\phi(^{7}\mathrm{Be})\over     \phi(^{7}\mathrm{Be})^\mathrm{SSM}}
  \right]^{2.08} 
\label{eq:s17} 
\end{eqnarray}
 Note    that   the   simple    exponent   0.5    on   the    S-factor   ratio
 $[\mathrm{S}_{11}/\mathrm{S}_{33}]$ is not  accidental, but reflects the fact
 that the $^3$He abundance has achieved  equilibrium in the region of the core
 where $^7$Be  and $^8$B neutrinos  are being produced.  The  number densities
 for $^3$He and protons are then related by
\begin{equation}
\left[ {N_3  \over N_p} \right]_\mathrm{equil}  = \sqrt{ \lambda_{pp}  \over 2
  \lambda_{33}} 
\end{equation}
where $\lambda_{pp}$  and $\lambda_{33}$ are  the local rates  proportional to
the  respective  S-factors.    Effectively  Eq.   (\ref{eq:s17})  states  that
laboratory  uncertainties in  S$_{17}$  (currently 7.5\%)  can  be traded  off
against  those in  S$_{34}$ (5.4\%)  and $\phi(^7$Be),  the most  poorly known
quantities  on  the  right-hand   side.   (Remember  that  all  S-factors  are
normalized to their  SFII best values.)  Adding errors  in quadrature, we find
that  this alternative  determination  yields $\mathrm{S}_{17}  = 0.967(1  \pm
0.117)$.  The  result will  not be competitive,  given the  current laboratory
precision of 7.5\%, unless the uncertainties on both S$_{34}$ and $\phi(^7$Be)
are reduced to $\sim$3\%.

{\it Summary:}  We have  refined the previous  arguments of \cite{HS}  to show
that  future  $^{15}$O  neutrino  flux  measurements  have  the  potential  to
constrain  the primordial core  metallicity of  C+N to  an accuracy  of $\sim$
10\%.   This would be  a very  significant result,  given that  differences in
recent  abundance  determinations  exceed   30\%.   The  method  exploits  the
additional  linear dependence  on  metallicity  of CN  cycle  burning, and  is
limited primarily  by expected uncertainties  of future experiments  like SNO+
and  by current  uncertainties  in laboratory  measurements  of nuclear  cross
sections.   The  nonnuclear  uncertainties  in the  relationship  we  derived,
previously  determined in  Monte  Carlo studies  to  be less  than $\sim$  3\%
\cite{HS},  are in  fact negligible  apart from  one parameter,  the diffusion
coefficient.  The dependence on diffusion is natural, reflecting the fact that
contemporary neutrino flux measurements are being used to constrain primordial
abundances, not  present-day core abundances.   Our primary test  of diffusion
and its  uncertainties comes from helioseismology, which  provided the initial
motivation for including  He and heavy-element diffusion in  solar models.  We
also point  out the possibility -- speculative  experimentally, but intriguing
theoretically  -- that  by also  measuring the  $^{13}$N solar  neutrinos, one
could  determine the separate  core abundances  of C  and N.   The present-day
burning of primordial C in the cooler outer core of the Sun contributes to the
$^{13}$N solar neutrino flux.

The idea behind the metallicity extraction is a general one: forming ratios of
neutrino fluxes that minimize the  sensitivity to core temperature and thus to
solar  model  uncertainties.   We  developed   a  second  example  of  such  a
minimum-uncertainty  SSM ratio  -- a  comparison  of $^7$Be  and scaled  $^8$B
neutrino  fluxes  --  that  isolates   a  specific  ratio  of  S-factors.   We
demonstrated  that the  precision to  which this  ratio is  known  from direct
laboratory  measurements is  in  fact identical  to  the precision  it can  be
determined from  measured solar  neutrino fluxes and  the SSM,  given existing
uncertainties  on nonnuclear input  parameters to  that model.   Thus neutrino
flux measurements have now  reached the precision where meaningful consistency
tests  with  laboratory  cross  sections  can  be done.   In  the  example  we
developed, the laboratory cross-section  measurements and neutrino fluxes were
found to be in excellent agreement.

\begin{acknowledgments}

AMS is  partially supported by the European  Union International Reintegration
Grant  PIRG-GA-2009-247732,  the  MICINN   grant  AYA2011-24704,  by  the  ESF
EUROCORES Program  EuroGENESIS (MICINN grant EUI2009-04170), by  SGR grants of
the Generalitat de Catalunya and by the EU-FEDER funds.
C.P-G  is supported  in  part  by the  Spanish  MICINN grants  FPA-2007-60323,
FPA2011-29678, the Generalitat Valenciana  grant PROMETEO/2009/116 and the ITN
INVISIBLES  (Marie   Curie  Actions,  PITN-GA-2011-289442).    This  work  was
supported  in  part   by  the  US  DOE  under   DE-SC00046548  (Berkeley)  and
DE-AC02-98CH10886  (LBL).  WH  thanks the  INT and  GSI for  their hospitality
while part  of this work was  done, and the Alexander  von Humboldt Foundation
for its support.
\end{acknowledgments}

\end{document}